\documentclass[%
 aip,
 amsmath,amssymb,
 reprint,%
]{revtex4-1}

\usepackage{graphicx}% Include figure files
\usepackage{dcolumn}% Align table columns on decimal point
\usepackage{bm}% bold math
\usepackage[utf8]{inputenc}
\usepackage[T1]{fontenc}
\usepackage{mathptmx}
\usepackage{etoolbox}

\makeatletter
\def\@email#1#2{%
 \endgroup
 \patchcmd{\titleblock@produce}
  {\frontmatter@RRAPformat}
  {\frontmatter@RRAPformat{\produce@RRAP{*#1\href{mailto:#2}{#2}}}\frontmatter@RRAPformat}
  {}{}
}%
\makeatother
\begin{document}

\preprint{AIP/123-QED}

\title{Preserving Coulomb blockade in transport spectroscopy of quantum dots, by dynamical tunnel-barrier compensation}
% Force line breaks with \\
\author{V. Jangir$^{\dagger}$}
\affiliation{Department of Physics, Indian Institute of Technology Bombay, INDIA}
\affiliation{Centre of Excellence in Quantum Information, Computing, Science \& Technology, IIT Bombay, INDIA}

\author{D. Shah$^{\dagger}$}
\affiliation{Department of Physics, Indian Institute of Technology Bombay, INDIA}

\author{S. Samanta}
\affiliation{Department of Physics, Indian Institute of Technology Bombay, INDIA}
\affiliation{Centre of Excellence in Quantum Information, Computing, Science \& Technology, IIT Bombay, INDIA}

\author{S. Rastogi}
\affiliation{Department of Physics, Indian Institute of Technology Bombay, INDIA}
\affiliation{Centre of Excellence in Quantum Information, Computing, Science \& Technology, IIT Bombay, INDIA}

\author{H.E. Beere}
\affiliation{Semiconductor Physics Group, Cavendish Laboratory, University of Cambridge, UK}

\author{D.A. Ritchie}
\affiliation{Semiconductor Physics Group, Cavendish Laboratory, University of Cambridge, UK}

\author{K. Das Gupta}
\affiliation{Department of Physics, Indian Institute of Technology Bombay, INDIA}
\affiliation{Centre of Excellence in Quantum Information, Computing, Science \& Technology, IIT Bombay, INDIA}

\author{S. Mahapatra}
\affiliation{Department of Physics, Indian Institute of Technology Bombay, INDIA}
\affiliation{Centre of Excellence in Quantum Information, Computing, Science \& Technology, IIT Bombay, INDIA}

\date{\today}% It is always \today, today,
\email{D. Shah: devashish\_shah@iitb.ac.in and S. Mahapatra: suddho@phy.iitb.ac.in}

\begin{abstract}
Surface-gated quantum dots (QDs) in semiconductor heterostructures represent a highly attractive platform for quantum computation and simulation. However, in this implementation, the barriers through which the QD is tunnel-coupled to source and drain reservoirs (or neighboring QDs) are usually non-rigid, and capacitively influenced by the plunger gate voltage ($V_P$). In transport spectroscopy measurements, this leads to complete suppression of current and lifting of Coulomb blockade, for large negative and positive values of $V_P$, respectively. Consequently, the charge-occupancy of the QD can be tuned over a rather small range of $V_P$. By dynamically tuning the tunnel barriers to compensate for the capacitive effect of $V_P$, here we demonstrate a protocol which allows the Coulomb blockade to be preserved over a remarkably large span of charge-occupancies, as demonstrated by clean Coulomb diamonds and well-resolved excited state features. The protocol will be highly beneficial for automated tuning and identification of the gate-voltage-space for optimal operation of QDs, in large arrays required for a scalable spin quantum computing architecture. \\

$\dagger$ These authors contributed equally to this work.
\end{abstract}

\maketitle
Electrostatically-induced quantum dots (QDs) in semiconductor heterostructures serve as the fundamental building blocks of the spin quantum computing architecture,\cite{Loss1998,Petta2005,Li2018,Fujita2017} and more recently, as a promising platform for quantum simulations\cite{Dehollain2020,vanDiepen2021_1,vanDiepen2021_2} of emergent many-body systems.\cite{Mourik2012,tenHaaf2024} For both applications, characterizing the QDs by Coulomb blockade spectroscopy (CBS) is an important initial step, to determine the energy states, electron g-factor, valley splitting, and coupling of the QD to the source/drain (S/D) reservoirs and the control gates.\cite{Tarucha1996,Nakaoka2004,Fuechsle2010} However, with fixed bias voltages applied to the barrier gates defining the QDs, obtaining high-quality CBS data over a sufficiently-large range of QD-occupancies is non-trivial. This is due to the fact that the plunger-gate voltage ($V_P$), which is ideally meant to control only the electrochemical potential of the QD, modifies the tunnel barriers between the QD and the S/D reservoirs. For increasing negative values of $V_P$, the tunnel-barrier transparency is suppressed, eventually rendering the source/drain current ($I_{SD}$) immeasurably low. On the other hand, for increasing positive values of $V_P$, the barrier transparency is enhanced, leading initially to cotunnelling and broadening of the excited-state fingerprints, and ultimately to complete lifting of the Coulomb blockade.\cite{Amasha2011}

In this work, we develop a protocol for recording the CBS data, wherein the tunnel-coupling of the QD to the S/D reservoirs is maintained constant across a wide range of plunger-gate voltages. By dynamically tuning the barrier-gate voltages to compensate for the effect of changing $V_P$,  we demonstrate the possibility of obtaining clean Coulomb diamonds and well-resolved excited states, for a large range of QD-occupancies, while also providing a systematic way to reach low electron numbers.

\begin{figure}[h!]
    \centering
    \includegraphics[width=\linewidth]{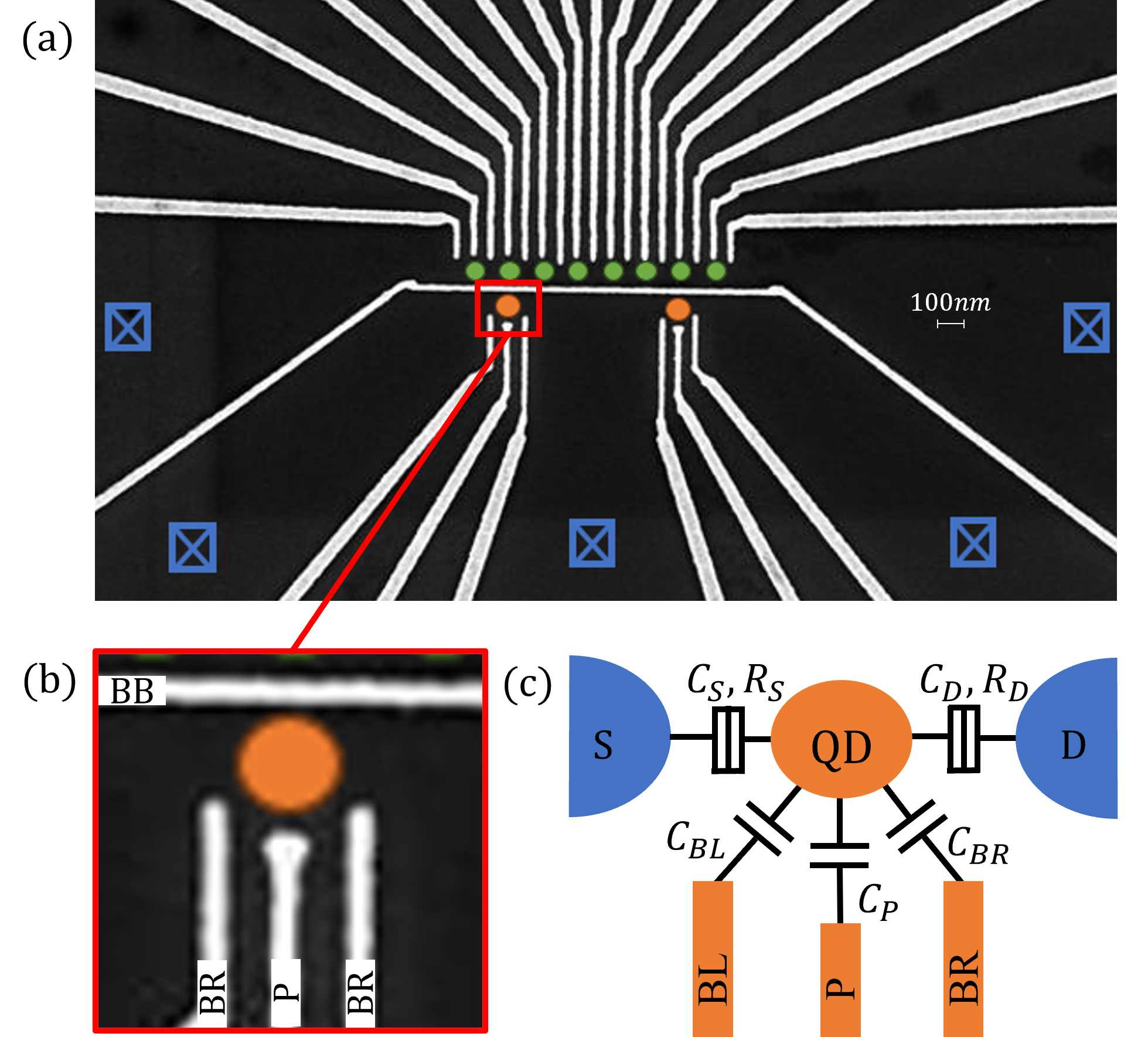}
    \caption{\textbf{A linear array of surface-gated QDs in GaAs/AlGaAs:} SEM image of a device equivalent to the one used for measurements, with a linear array of eight QDs and two SET charge sensors. (b) Close-up view of (a), showing the barrier ($BL, BR$) and plunger $(P)$ gates of the SET, studied here. (c) Schematic representation of the equivalent circuit of the SET, showing capacitive couplings of the QD to the barrier gates $(C_{BL/R})$ and with the plunger gate $(C_P)$, and tunnel coupling $[C_{S(D)}, R_{S(D)}]$ to the source (drain) reservoir.}
    \label{fig:SEM_and_schematic}
\end{figure}
\begin{figure*}[t]
    \centering
    \includegraphics[width=0.95\linewidth]{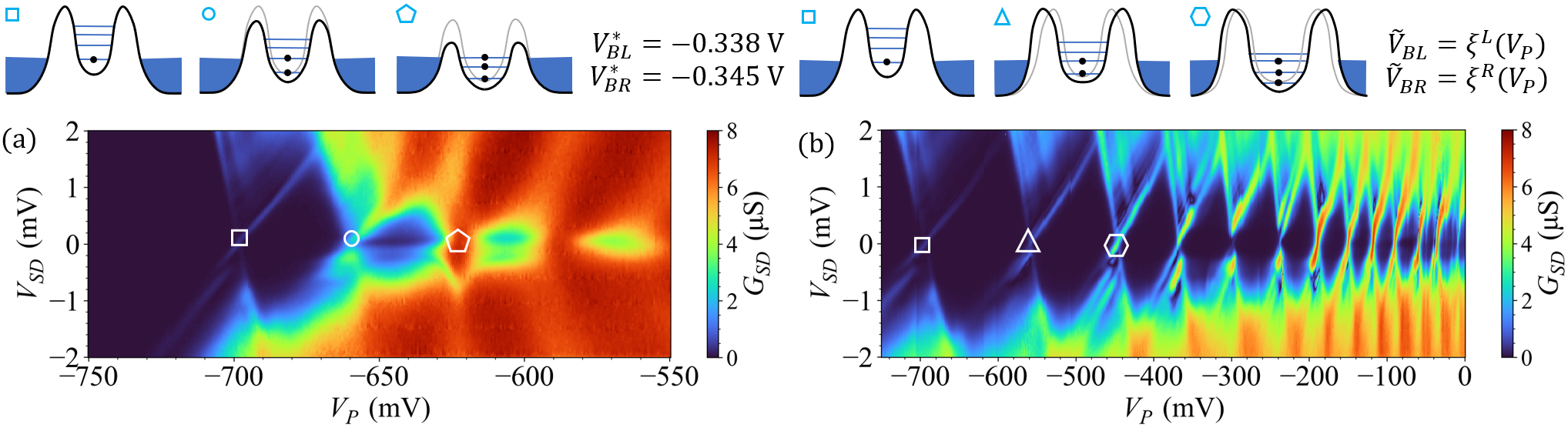}
    \caption{\textbf{Coulomb Blockade Spectroscopy (CBS)}: Conductance color-map obtained using (a) the conventional and (b) the compensated techniques of recording CBS data. The schematics on top depict how the plunger-gate voltage affects the tunnel-barriers in either case. In the case of the compensated data, dynamical tuning of barrier-gate voltages allows the barrier transparencies to be maintained approximately constant. Both measurements have identical barrier-gate voltages when $v_P=-700\,\mathrm{mV}.$ }
    \label{fig:Diamonds_Schematic}
\end{figure*}

To demonstrate the measurement protocol, we use QDs defined in a modulation-doped GaAs/AlGaAs heterostructure, consisting of a two-dimensional electron gas (2DEG) formed at the heterojunction, with an electron concentration and mobility of $1.61\times 10^{11} \,\mathrm{cm}^{-2}$ and $9.78 \times 10^5 \,\mathrm{cm^2/Vs}$ (at $4 \,\mathrm{K}$), respectively. The plan-view SEM image of the surface gates, defining an array of eight QDs, and two single-electron-transistor (SET) charge sensors, is shown in Fig.\ref{fig:SEM_and_schematic}(a). For this work,  we recorded the CBS data for the left SET (enlarged in Fig.\ref{fig:SEM_and_schematic} (b)), consisting of a relatively large dot with a lithographic size of 200 nm. Fig.\ref{fig:SEM_and_schematic}(c) schematically shows the equivalent circuit diagram of the SET, depicting the tunnel coupling of the QD to the source/drain reservoirs (S/D) and the capacitive coupling to the barrier gates (BL and BR) and the plunger gate (P). While the narrow surface gates (50 nm wide and separated by 75 nm) allow strong confinement of electrons  and tunable tunnel-coupling of the QD to the S/D reservoirs, complex iterative techniques are required to observe well-resolved Coulomb blockade oscillations, in the usual approach for tuning the QD.

Fig.\ref{fig:Diamonds_Schematic}(a) plots the differential conductance of the SET sensor, as a function of $V_P$ and the source-drain bias ($V_{SD}$), recorded by the conventional approach (i.e. with a fixed bias on the barrier gates BL and BR). As observed, the Coulomb diamonds are barely defined for a change in the QD occupancy beyond two electrons. Strong inelastic co-tunneling is observed to set in first, followed by complete lifting of the blockade, when more than four additional electrons are loaded onto the QD. On the other hand, the power of the protocol developed in this work is evident in the CBS data depicted in Fig. \ref{fig:Diamonds_Schematic}(b). The differential conductance recorded with dynamical control of the tunnel-barrier transparencies reveals clean Coulomb diamonds, with well resolved excited-state features, for QD-occupancy change of up to thirteen electrons. The persistence of robust Coulomb blockade for such a large range of plunger gate voltages has not been demonstrated before, for surface-gated QDs in (modulation-doped) semiconductor heterostructures. 

\begin{figure}[h!]
    \centering
    \includegraphics[width=\linewidth]{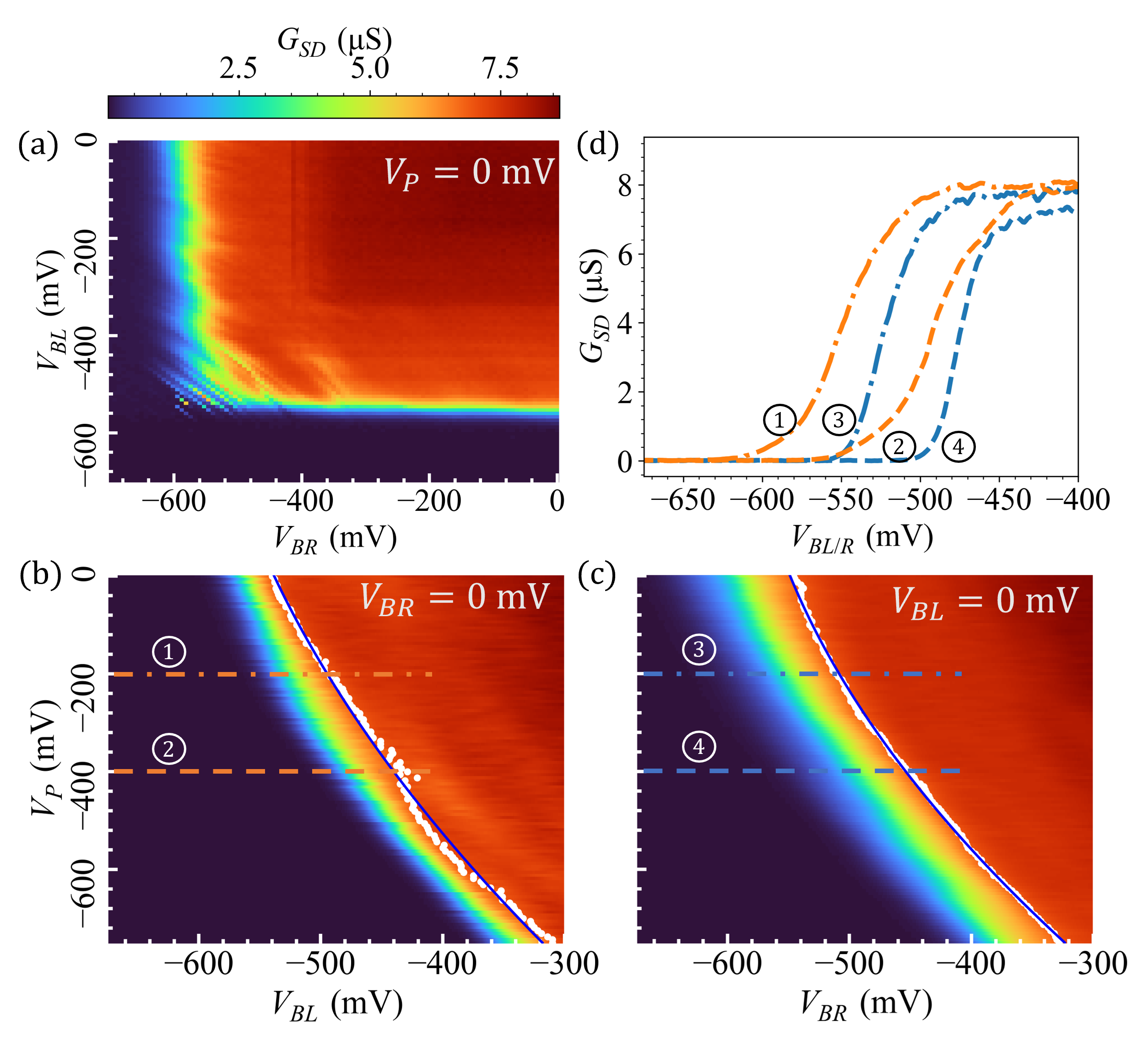}
    \caption{\textbf{Quantifying the effect of $V_P$ on tunnel barrier transparency:} (a) Conductance-map recorded as a function of the barrier-gate voltages, $V_{BL}$  and $V_{BR}$. Conductance oscillations at the “corner" of this plot indicate the onset of Coulomb blockade. (b), (c) Conductance maps recorded over the $V_{BR}$ – $V_P$ and $V_{BL}$-$V_P$ gate-voltage spaces, where $V_P$ denotes the plunger-gate voltage. The capacitive effect of the plunger-gate voltage on the tunnel-barrier transparency causes a shift in the barrier pinch-off voltages. (d) $G_{SD}$ versus $V_{BL/BR}$ plots for two different values of $V_P$ explicitly show the shift in pinch-off voltages due to capacitive effect of the latter, on the barrier-transparency. }
    \label{fig:Corners}
\end{figure}

The underlying phenomenological difference between the conventional and the current approach is illustrated in the schematic representation of the SET energy landscape, in Figs. 2(a) and 2(b). For the conventional approach, the tunnel-barriers are strongly suppressed with increasing $V_P$, while they are nearly unaltered, when the dynamical control protocol is used. We emphasize here that the protocol does not merely preserve the visibility of the Coulomb diamonds, but actually ensures persistence of the Coulomb blockade phenomenon, for a large range of QD occupancies.

We now detail the protocol of dynamical control of the tunnel-barrier transparency, developed here. To record CBS data in the usual approach, the QD is first "tuned" to the Coulomb blockade regime.\cite{Beenakker1991} The SET conductance ($G_{SD}$) as a function of the barrier gate voltages ($V_{BL}$ and $V_{BR}$) is shown in Fig. \ref{fig:Corners}(a), recorded for a fixed plunger gate voltage ($V_P=0$) and a small source-drain bias ($V_{SD}=100\,\mathrm{\mu V}$). The Coulomb blockade regime is identified by the appearance of oscillatory $G_{SD}$ in the color-map, which represent conductance peaks, separated by the blockaded regions. This measurement is repeated for different $V_P$ values. Fixed values of $V_{BL}$ and $V_{BR}$ are then chosen from the map with the best-resolved Coulomb oscillations, and $V_P$ is swept to record the CBS data. For these fixed values of $V_{BL}$ and $V_{BR}$, the tunnel rates $\Gamma_L$ and $\Gamma_R$  vary widely over the range of $V_P$, typically spanned in a CBS measurement. To maintain $\Gamma_L$ and $\Gamma_R$ nearly constant, as $V_P$ is varied, we dynamically adjust the barrier-gate voltages such that: 
\begin{equation}
    \tilde{V}_{BL(R)} = V^0_{BL(R)} + \xi^{L(R)}(V_P).
\end{equation}
Here, $V^0_{BL(R)}$ is the left (right) barrier-gate voltage that gives a desired tunneling conductance at $V_P = 0$, while $\xi^{L(R)}$ is a polynomial function of $V_P$ that maintains this conductance, when $V_P$ is swept to record the CBS data. To determine $\xi$, we additionally record conductance maps on the $V_{BL}-V_P$ ($V_{BR}-V_P$) gate-space, while maintaining $V_{BR}$ = 0 ($V_{BL}$ = 0), as shown in Fig. \ref{fig:Corners}(b) (Fig. \ref{fig:Corners}(c)). It is clearly evident that the onset of $G_{SD}$ in both color-maps depends strongly on $V_P$. The line-plots in Fig. \ref{fig:Corners}(d) explicitly show the shift in the pinch-off regime (where $G_{SD}$ drops from $\sim 80\%-20\% $ of its maximum value) of both $V_{BL}$ and $V_{BR}$, when $V_P$ is reduced from -200 $\mathrm{mV}$ to -400 $\mathrm{mV}$.  

In the pinch-off regime, we identify constant-conductance contours, as shown by the white dotted curves in Figs. \ref{fig:Corners}(b) and \ref{fig:Corners}(c). These contours with fixed conductance, $G_0$, are fitted with second-order polynomial functions, such that

\begin{equation}
    \tilde{V}_{BL(R)} = V^0_{BL(R)}+\alpha_{L(R)} V_P + \beta_{L(R)} V_P^2.
    \label{eq: fit}
\end{equation} 

When $V_{BL}, V_{BR}$ are simultaneously tuned according to Eq. (\ref{eq: fit}), while sweeping the $V_P$ to change the QD-occupancy, we obtain the uniformly-resolved CBS data, as shown in Fig. \ref{fig:Diamonds_Schematic}(b). Fig. \ref{fig:Corners}(a) reveals that the pinch-off regime of $V_{BL}$ is almost independent of $V_{BR}$, and vice-versa, in contrast to the influence of $V_P$ on either of them. Thus, we ignore the effect of $V_{BL}$ ($V_{BR}$) on the right (left)-barrier tunneling-rate. Also, higher order terms in $\xi$, $\mathcal{O}(V_P^3, V_P^4, ...)$, are found to be negligible compared to $\beta V_P^2$, and are therefore neglected.

\begin{figure}[h]
    \centering
    \includegraphics[width=\linewidth]{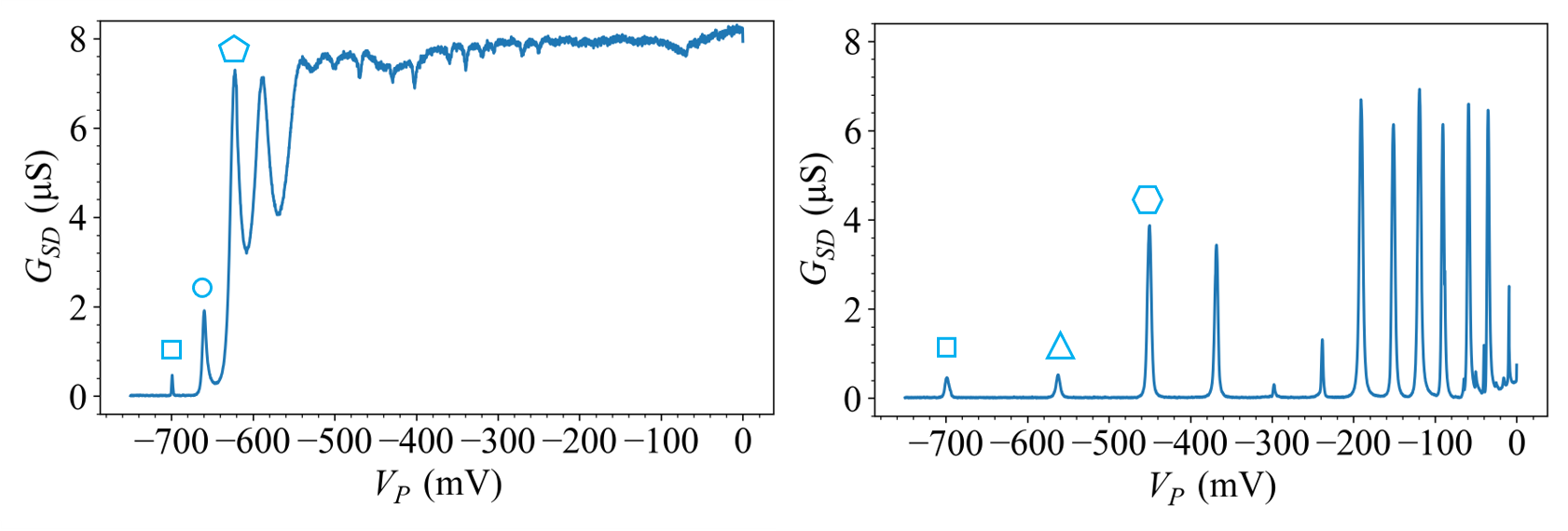}
    \caption{\textbf{Coulomb oscillations at low bias:} Plot of the SET conductance measured (a) without and (b) with barrier-gate voltage compensation, with varying $V_P$, for $V_{SD} = 100 \mathrm{\mu V}$. For both cases, barrier voltages $V_{BL}^*=\tilde{V}_{BL}=-0.338\,\mathrm{V}$ and $V_{BR}^* =\tilde{V}_{BR}= 0.345\,\mathrm{V}$ are identical when $V_P = -700\,\mathrm{mV}$.}
    \label{fig:Coulomb_1D}
\end{figure}

Within this protocol of recording the CBS data, only a single parameter, namely $G_0$, must be chosen from the pinch-off region. Fig. \ref{fig:Coulomb_1D} compares the low-bias ($V_{SD}= 100\,\mathrm{\mu V}$) conductance plots recorded by the conventional approach (Fig. \ref{fig:Coulomb_1D}(a)) and by dynamically compensating the barrier-gate voltages (Fig. \ref{fig:Coulomb_1D}(b)). The parameters for the measurement are listed in Table. \ref{Table}. Sharp Coulomb peaks, separated by well-defined blockaded regions are clearly visible in the latter case, in contrast to that in the former. 
\begin{table}[h]
\caption{\label{tab:compare} Parameters used for recording  Coulomb oscillations, with and without barrier-gate voltage compensation}
\begin{ruledtabular}
\begin{tabular}{cc}
Uncompensated&Compensated\\
\hline
$\Gamma_{L(R)}\neq \mathrm{constant}$ & $\Gamma_{L(R)}\simeq \mathrm{constant}$\\
% $T^{S(D)}\neq \mathrm{constant}$ & $T^{S(D)}\simeq \mathrm{constant}$\\
$V^*_{BL} = -0.338\mathrm{V}$ & 
$\tilde{V}_{BL} = -0.538 -0.190 V_P + 0.138V_P^2$\\
$V^*_{BR} = -0.345\mathrm{V}$&
$\tilde{V}_{BR} = -0.549 -0.165 V_P + 0.181V_P^2$\\
\end{tabular}
\end{ruledtabular}
\label{Table}
\end{table}

\begin{figure}[t]
    \centering
    \includegraphics[width=\linewidth]{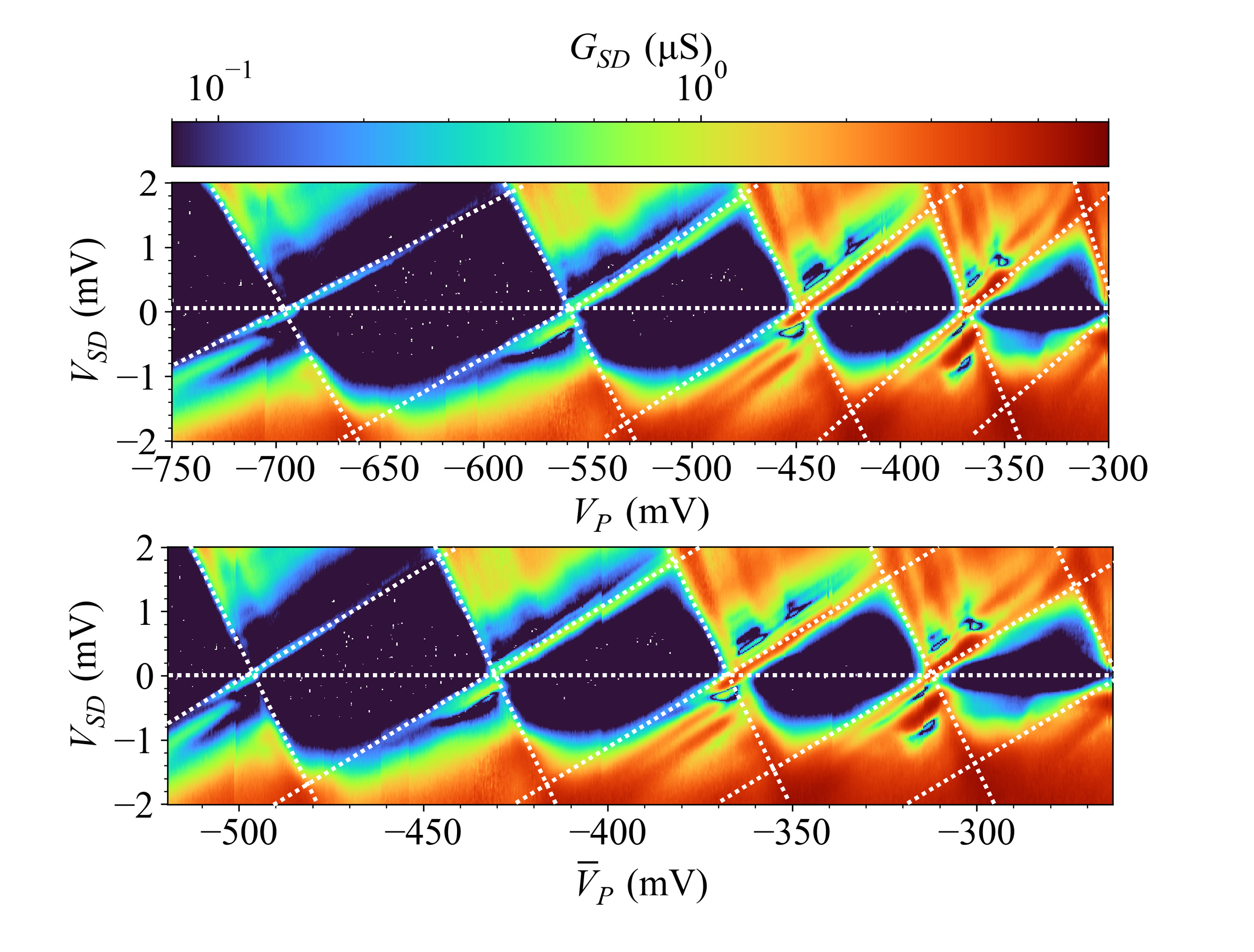}
    \caption{\textbf{Coulomb blockade maps before and after rescaling}. Close-up view of the last four diamonds for the compensated sweep (a) without and (b) with rescaling of the plunger-gate-voltage axis. The rescaling process makes the effective level arm independent of the (rescaled) plunger-gate voltage, and thereby, restores the uniformity of the edges bounding the Coulomb diamonds.}
    \label{fig:Rescaling}
\end{figure}

Unlike in the conventional approach, the Coulomb diamonds obtained in the compensated CBS data are bounded by edges with non-uniform slopes (Fig. \ref{fig:Rescaling}(a)). This occurs due to the quadratic nature of the compensation voltages, $\tilde{V}_{BL/R}$, applied to the barrier gates, leading to a $V_P$-dependent ``effective lever arm". A specific re-scaling of the plunger-gate voltage ($V_P\rightarrow \bar{V}_P$) allows the linear dependence of QD-electrochemical-potential on $\bar{V}_P$, and hence the uniform slopes of the Coulomb-diamond edges, to be restored. For the device in Fig. \ref{fig:SEM_and_schematic}c, we modify the usual expression of the electrochemical potential of the $N$-electron QD, $\mu_{QD} (N)$,  to explicitly include the dependence on $V_{BL}$ and $V_{BR}$ as:
\begin{equation}
\label{Eq: mu dot}
\begin{split}
    \mu_{QD}(N) = &\: \frac{e^2}{C_\Sigma}\left(N-N_0-\frac{1}{2}\right)-\frac{e}{C_\Sigma}(C_{BL}V_{BL}+ C_{BR}V_{BR}) \\ & -\frac{e}{C_\Sigma}(C_PV_P + C_SV_S + C_DV_D)  + E_N.
\end{split}
\end{equation}
Here, $C_i$ denote the mutual capacitances as shown in Fig.\ref{fig:SEM_and_schematic}(c) and $V_S$ and $V_D$, the voltages on the source and drain contacts, respectively. $C_\Sigma = C_S + C_D + C_P + C_{BL} + C_{BR}$ is the summed capacitance, $N_0$, the background charge, and $E_N$, the energy eigenstate of the N-electron ground state of the QD. Since barrier-gate voltages are kept constant in the conventional approach, the terms involving $C_{BL}$ and $C_{BR}$ are included in $N_0$, and are ignored (relative to $C_{S} + C_{D}$) when calculating the summed capacitance $C_\Sigma$. With dynamical control of the barrier-gate voltages on the other hand, these terms need to be considered explicitly. For asymmetric S/D biasing ($V_S=V_{SD}, V_D=0$) the two edges bounding the Coulomb diamond from the left are obtained by equating $\mu_{S} (=-eV_s)$ and $\mu_{D} (= 0)$ to $\mu_{QD}(N)$, which may be written as: 

\begin{equation}
\label{Eq: mu dot expanded}
\begin{split}
    \mu_{QD}(N) = g(N) - \frac{e}{C_\Sigma}(\bar{C}_PV_P + \kappa_PV_P^2+ C_SV_S).
    \end{split}
\end{equation}

Here, $g(N) = e^2(N-N_0-0.5)/C_\Sigma + e(C_{BL}V^0_{BL}+ C_{BR}V^0_{BR})/C_\Sigma$ is a constant independent of $V_P$ and $V_{SD}$. The combined capacitance $\bar{C}_P = C_P + \alpha_LC_{BL} + \alpha_RC_{BR}$ and $\kappa_P = \beta_LC_{BL} + \beta_RC_{BR}$ are also constants. Using Eq. (\ref{Eq: mu dot expanded}), and the analogous expression for the $N+1$ occupancy, it can be shown that:
\begin{equation}
    \Delta V_P^{(N)} = \frac{\Delta V_{SD}^{(N)} C_\Sigma}{\bar{C}_P + \kappa_P\left(V_P^{0(N-1,N)}+V_P^{0(N,N+1)}\right)} = \alpha_P(V_P)\Delta V_{SD}^{(N)}.
\end{equation}
Here, $\Delta V_{SD}^N$ is the height and $\Delta V_P^N = \left|V_P^{0(N-1,N)}-V_P^{0(N,N+1)}\right|$ is the width of the $N$-occupancy diamond, where $V_P^{0(i-1,i)}$ are the plunger voltages for which $eV_{SD} = \mu_{QD}(i) = 0$. The effective lever arm $\alpha_P$ is a gate voltage-dependent quantity. Using a two-variable least square fitting for the above measurables for the last five diamonds, $\kappa/\bar{C}_P$ is found to be $0.41\mathrm{V^{-1}}$. Rescaling the plunger-gate-voltage axis by defining $\bar{V}_P = V_P(1+0.41V_P)$ gives:
\begin{equation}
\begin{split}
    \mu_{QD}(N) &= g(N) - \frac{e}{C_\Sigma}(\bar{C}_P\bar{V}_P+C_SV_S)\\&=g(N)-e\bar{\alpha}_P\bar{V}_P-e\alpha_S V_S.
\end{split}
\end{equation}
This restores the expected constant slope of the diamonds (Fig. \ref{fig:Rescaling}(b)), simplifying any further data analysis. More importantly, varying $\bar{V}_P$ leads to a proportional linear change of the electrochemical potential while ensuring that $C_{S/D}$ and $\Gamma_{S/D}$ are maintained nearly constant. However, changes in $C_P$ caused by a change in the shape and location of the QD, due to local charge rearrangement at specific values of $V_P$, cannot be accounted for, within this approach.

In conclusion, we reported the development of a protocol for acquisition of high-quality Coulomb blockade data, employing dynamical compensation of the voltages defining the barriers of the probed quantum dot. By this technique, Coulomb blockade is preserved for a large span of plunger-gate voltages, thus allowing clean Coulomb diamonds, with well-resolved excited state features, to be recorded for a range of electron occupancies, which is otherwise inaccessible by conventional techniques. Applications in physics-informed automated tuning of quantum dots in large arrays and identification of optimal operation regimes in the gate-voltage space \cite{Kalantre2019,Ziegler2023, Volk2019, Hsiao2020} will greatly benefit from the protocol developed in this work. Another direct application of this work is in tuning charge-sensor dots, wherein it is critical to obtain narrow Coulomb peaks, and thereby maximize the charge contrast.\cite{Berkovits2005, Mahapatra2011} 

\section*{Methods}
 The device is fabricated on a modulation-doped GaAs$/$AlGaAs heterostructure, with a quantum well $90\, \mathrm{nm}$ below the surface. The 2D electron gas has a concentration of $1.61\times 10^{11} \,\mathrm{cm}^{-2}$ and mobility of $9.78 \times 10^5 \,\mathrm{cm^2/Vs}$ (at $4 \,\mathrm{K}$). The heterostructure was grown by molecular beam epitaxy at the Cavendish Laboratory of Cambridge University, while all device fabrication steps were carried out at the IIT Bombay Nanofabrication Facility. A mesa is created using photolithography and wet-chemical etching as the first step of device fabrication. The design consists of five ohmic contacts connected to the 2DEG within the mesa region. Approximately $50\,-\mathrm{nm}$- wide depletion gates are patterned using electron-beam lithography and electron-beam evaporation for deposition of a Ti/Au metal stack, followed by a standard lift-off process. All transport measurements were conducted in a Bluefors LD400 refrigerator with a base temperature of $10\,\mathrm{mK}$. AC measurements were done using a low-frequency lock-in amplifier.

\section*{Author Contributions}

V. Jangir: fabrication (lead); measurement (supporting); analysis (equal); writing – original draft (lead),
D. Shah: conceptualization (equal); measurement (lead); analysis (equal); writing – original draft (lead),
S. Samanta: measurement (lead); analysis (equal); writing – original draft (supporting),
S. Rastogi: fabrication (lead); measurement (supporting); writing – original draft (supporting),
H.E. Beere: sample preparation (lead), D.A. Ritchie: sample preparation (lead),
K. Dasgupta: sample preparation (supporting); measurement (supporting),
S. Mahapatra: writing – review and editing (lead); conceptualization (equal); analysis (equal); measurement (supporting); fabrication (supporting); supervision (lead).\\

The authors declare no conflict of interest.

\section*{Data Availability Statement}
The data that support the findings of this study are available from the corresponding authors upon reasonable request.

\begin{acknowledgments}
We acknowledge the insightful discussions with Uditendu Mukhopadhyay, Department of Physics, IIT Bombay. We thank the Defence Research and Development Organization (DRDO) and the Department of Science and Technology (DST) of the Ministry of Education (MoE), Government of India (GoI), for the funding support and the IIT Bombay Nanofabrication Facility (IITBNF) management and staff for support with device fabrication.
\end{acknowledgments} 

\section*{Citations and References}

\end{document}